\documentclass[a4paper,preprint,eqsecnum,showpacs,prb]{revtex4}
\usepackage[T1]{fontenc}
\usepackage[latin1]{inputenc}
\usepackage{graphicx}
\usepackage{amssymb}


\begin{document}

\title{Spectral analysis of resonant x-ray scattering in CeB$_6$\\ 
under an external magnetic field
}

\author{Tatsuya Nagao$^1$ and Jun-ichi Igarashi$^2$}

\affiliation{$^1$Faculty of Engineering, Gunma University, Kiryu, 
Gunma 376-8515, Japan \\
$^2$Faculty of Science, Ibaraki University, Mito,
Ibaraki 310-8512, Japan
}

\begin{abstract}
We study the resonant x-ray scattering (RXS) spectra of CeB$_{6}$
in an antiferroquadrupole (AFQ) ordering phase,
near the Ce $L_3$ edge under the applied magnetic field $\textbf{H}
\parallel (\overline{1},1,0)$. 
On the basis of a localized electron model equipped with a mechanism 
that the RXS signal is brought about by the
intra-atomic Coulomb interaction in Ce, 
we calculate the RXS spectra.
The obtained spectra exhibit two contributions
around the electric dipole ($E$1) and quadrupole ($E$2) positions, 
and differ drastically when
the orientation of $\textbf{H}$ is reversed.
The difference is brought about by the cross terms between the even-rank 
AFQ and magnetic-induced odd-rank contributions.
At the $E$1 region, the relevant cross term is the one
within the $E$1 process, while at the $E$2 region,
they are ones within the $E$2 process and between the $E$1 and $E$2 processes.
These findings capture the characteristic features
the recent experimental data show, and provide a strong support and information
of the field-induced multipole orderings.
We also evaluate the RXS spectra near the $L_{2}$ edge.
Though the results show no $E$2 contribution, we find that
the intensity around the $E$1 transition, which is as large as
that at the $L_{3}$ edge, can be detected experimentally.

\end{abstract}

\pacs{75.10.-b, 75.25.Dk, 78.70.Ck,78.20.Bh}

\maketitle

\section{Introduction\label{sect.1}}
Numerous and intensive research activities have been concentrated
in the field of the $f$-electron systems where mutual interplay
of charge, spin, and orbital degrees of freedom 
produces a rich variety of physical properties
characterizing the strongly correlated 
electron systems. Among a large number of interests,
the materialization of the
multipolar ordering phase in these systems 
is one of the most fascinating topics 
to be addressed.\cite{Santini09,Kuramoto09}

Resonant x-ray scattering (RXS) is one of the most promising
probes to observe experimental evidences of the multipole
order with rank higher than two.
A tensorial nature of the scattering amplitude helps
our understanding of the experimental 
result.\cite{Hannon88,Blume94,Lovesey05,Dmitrienko05}
However, an analysis based merely on the symmetrical consideration of 
the tensorial nature does not answer the origin of the observed RXS signals.
For instance, when the RXS signals were detected at the Mn
$K$-edge in the orbital ordering phase of manganites.
\cite{Murakami98.1,Murakami98.2}
There was a controversy on what brought about the observed signals. 
In the electric dipole ($E$1) transition,
the $K$-edge signals reflect the anisotropic charge
distribution of the $4p$ states, which form bands. 
Such an anisotropy may be given rise to by the distortion of the lattice
or by the underlying ordering pattern of the $3d$ electrons through the
intra-atomic Coulomb interaction.\cite{Ishihara98}
We call the latter as the "\textit{Coulomb mechanism}".
Extensive investigations have revealed that the observed intensities were
originated from the lattice distortion.\cite{Elfimov99,Benfatto99,Takahashi99}
Since the $4p$ states of transition metals are rather extended in space,
this result is quite reasonable. It is now recognized that the same mechanism 
is working on the RXS in other transition-metal compounds such as YTiO$_3$ and 
YVO$_3$.\cite{DiMatteo09,Takahashi01,Takahashi02}

The situation is different for RXS near the $L$-edge in rare-earth compounds,
since the $4f$ states are so localized in space that the lattice distortion
associated with multipole ordering is expected to be much smaller than that 
in transition-metal compounds.
In CeB$_6$, the antiferroquadrupolar (AFQ) ordering phase is
inferred from various indirect observations
such as macroscopic measurements, resonance methods, neutron scattering, 
and so on.\cite{Takigawa83,Effantin85,Terzioglu01,Zaharko03,Schenck04,Plakhty05}
The most direct evidence of the AFQ order is provided by the RXS experiment
by Nakao \textit{et al}.\cite{Nakao01} and Yakhou \textit{et al}.,
\cite{Yakhou01} who have succeeded in detecting the RXS signal
at the Ce $L_3$ edge at an AFQ Bragg spot 
$\textbf{G}=\left(\frac{1}{2},\frac{1}{2},\frac{1}{2}\right)$.
Later, another experimental support was given by Tanaka \textit{et al}.
from the non-resonant x-ray Thomson scatteing (NRXTS) study which detected
directly an evidence of the aspherical charge density.\cite{Tanaka04}
Note that no lattice distortionis are 
observed in this material.\cite{Akatsu03,Sikora05} 

In our previous papers, we have calculated microscopically the RXS spectra 
on the basis of a localized electron picture which relies on the Coulomb 
mechanism, and have obtained the spectra in agreement with the experiments.
\cite{Nagao01,Igarashi02}
In addition, invoking the same theoretical framework elaborated in
Refs. \onlinecite{Nagao01} and \onlinecite{Igarashi02}, 
we have calculated the NRXTS intensities on the AFQ Bragg spots.
Our result has reproduced well the relative intensities
of RXS and NRXTS and a Fano-dip like structure
at $\textbf{G}=\left(\frac{5}{2},\frac{3}{2},\frac{3}{2}\right)$,
\cite{Nagao03}
in accordance with the Yakhou \textit{et al}.'s data.\cite{Yakhou01} 

It is predicted that the applied magnetic field induces the 
antiferro-octupole 
ordering in CeB$_6$,\cite{Shiina97,Shiina98} 
which could be detected by RXS 
and neutron scattering\cite{Kuwahara07} measurements.  
Matsumura \textit{et al}. have recently succeeded in detecting 
the octupole ordering induced by the applied magnetic field
via RXS.\cite{Matsumura09}
By using different experimental settings examined before,\cite{Nakao01,Yakhou01}
they observed the RXS spectra near the Ce $L_{3}$ edge at 
$\textbf{G}=\left(\frac{3}{2},\frac{3}{2},\frac{1}{2}\right)$
under the applied field along $\textbf{H} \parallel (\overline{1},1,0)$
in the AFQ phase. Their data illustrate three notable features.
First, in addition to main peak of the $E$1 transition
around 5724 eV, the spectra show a small peak of the electric
quadrupole ($E$2) transition around 5718 eV
when the field is in the plus direction ($+\textbf{H}$).
This finding makes a remarkable contrast with the previous RXS data
where the $E$2 peak is practically absent both experimentally and
theoretically.\cite{Nakao01,Yakhou01,Nagao01,Igarashi02}
Second, when the orientation of the applied field is reversed,
the spectral shape drastically changes.
That is, the small peak around the $E$2 position becomes
obscure when the field is in the minus direction ($-\textbf{H}$). 
Third, the peak intensity at the $E$1 peak varies in certain amount
when the direction of $\textbf{H}$ is reversed. 
Note that although
the final feature was not emphasized in Ref. \onlinecite{Matsumura09},
they confirmed that the difference actually exists,
in particular, for the field strength approximately
larger than 2 $T$.\cite{com.Matsumura}

These features were analyzed to originate from 
the cross terms between the contributions of 
primary AFQ (even rank) order parameters and those of field-induced
(odd rank) order parameters.
The purpose of this paper is to elucidate those observations
by analyzing quantitatively the spectra from a microscopic standpoint 
beyond qualitative one based merely on the symmetry consideration.
\cite{Lovesey02}
Developing our previous treatment,\cite{Nagao01,Igarashi02,Nagao03} 
we find that the field dependence comes from the interference effect 
between the even rank and odd rank signals.
Some of them come from the terms within the $E$1 process
and within the $E$2 process, while others come from the terms
between the $E$1 and $E$2 processes. 
Since the field-induced octupole ordering, for example, gives
rise to the third order rank signal,
the information of octupole ordering could be extracted from
the field dependence of the spectra.

For numerical calculation, we adopt the same model and parameter settings 
to the previous works as possible as we can to keep continuity of the research. 
We obtain the spectra reproducing semiquantitatively the three features 
the experiment had revealed.
Note that the `\emph{fast collision approximation}', which is widely used to
analyze the spectra, is insufficient to discuss the field dependence,
since it predicts no field dependence of the main peak intensity.
Finally, we find that the RXS spectrum at the Ce
$L_{2}$ edge has enough intensity to be detected experimentally.

This paper is organized as follows. In Sec. \ref{sect.2}, we 
introduce a theoretical framework to investigate
the RXS intensity in the localized electron systems. 
In Sec. \ref{sect.3}, we briefly summarize a model Hamiltonian
which describes the initial state and used in our previous works,
and explain the mean field solution of the Hamiltonian.
Also the intermediate states of the scattering processes are
presented.
Numerical results of the calculated 
RXS spectra are shown in Sec. \ref{sect.4}
with comparisons with the experimental results.
The last section is devoted to concluding remarks.

\section{Theoretical Framework \label{sect.2}}
\subsection{Scattering amplitude}
RXS is described as a second order optical process: photon with frequency 
$\omega$, wave number $\textbf{k}$, and polarization $\mu$ ($=\sigma$ or $\pi$)
is diffracted by the sample into the state with the same frequency $\omega$,
wave number $\textbf{k}'$, and polarization $\mu'$ ($=\sigma'$ or $\pi'$).
The amplitude is approximated by a sum of the contributions
from each Ce ion, which can be written as
\begin{equation}
F(\textbf{k},\mbox{\boldmath{$\epsilon$}};
\textbf{k}',\mbox{\boldmath{$\epsilon$}}':\omega)
= \sum_{n=1,2} f_{n} (\textbf{k},\mbox{\boldmath{$\epsilon$}};
\textbf{k}',\mbox{\boldmath{$\epsilon$}}':\omega),
\label{eq.amp.1}
\end{equation}
where $f_{n}$ stands for the scattering amplitude of the
$En$-$En$ transition and electric $2^{n}$-th pole process is 
abbreviated to $En$.
We omit the contributions from the terms like $En$-$Em$ 
with $n \neq m$
since we restrict our attention to centrosymmetric system.
In this context, the $En$-$En$ transition is simply called as the
$En$ transition hereafter.
The $En$ amplitude is written as
\begin{equation}
f_{n}(\textbf{k},\mbox{\boldmath{$\epsilon$}};
\textbf{k}',\mbox{\boldmath{$\epsilon$}}':\omega)
\propto \frac{1}{\sqrt{N}}
\sum_{j} \textrm{e}^{-i\textbf{G} \cdot \textbf{r}_j}
M_{n} (j:\textbf{k},\mbox{\boldmath{$\epsilon$}};
\textbf{k}',\mbox{\boldmath{$\epsilon$}}':\omega), \label{eq.amp.2}
\end{equation}
where $M_{n} (j:\textbf{k},\mbox{\boldmath{$\epsilon$}};
\textbf{k}',\mbox{\boldmath{$\epsilon$}}':\omega)$ represents
the RXS amplitude at site $j$ with position vector $\textbf{r}_{j}$.
The number of Ce site is denoted as $N$.
Scattering vector $\textbf{G}$ is defined by $\textbf{k}'-\textbf{k}$.
Note that the above expressions aimed for absolute zero temperature
are easily extended for
a treatment of finite temperature ($T$) case by multiplying
probability $p_{j,m}$ to 
$M_{n} (j:\textbf{k},\mbox{\boldmath{$\epsilon$}};
\textbf{k}',\mbox{\boldmath{$\epsilon$}}':\omega)$
and summing over $m$.
Here $p_{j,m}$ is proportional to the Boltzmann factor $\textrm{e}^{-E_{j,m}/T}$
with $E_{j,m}$ being the energy of the $m$-th eigenstate at site $j$.
For simplicity, we proceed the formulation for $T=0$,
though our numerical calculations in Sec. \ref{sect.4}
will be those for finite temperatures.

The $E$1 amplitude is 
\begin{equation}
M_{1} (j:\mbox{\boldmath{$\epsilon$}};\mbox{\boldmath{$\epsilon$}}':\omega)
=\sum_{\mu,\mu'=1}^{3} \epsilon_{\mu}' \epsilon_{\mu'}
\sum_{\Lambda}
\frac{\langle 0| x_{\mu,j} | \Lambda \rangle
\langle \Lambda | x_{\mu',j} | 0 \rangle}
{\hbar \omega - (E_{\Lambda}-E_0) + i \Gamma}, \label{eq.def.M1}
\end{equation}
where $|0 \rangle$ denotes the ground state
with eigenenergy $E_0$, while $|\Lambda \rangle$
denotes the intermediate state with eigenenergy $E_{\Lambda}$.
The lifetime broadening width of the core
hole is represented by $\Gamma$ and it is fixed to be 1.5 eV
in this work. The dipole operators $x_{\mu,j}$ are
described as $x_{j}, y_{j}$, and $z_{j}$ for
$\mu=1, 2$, and $3$, respectively, in the coordinate
system fixed to the crystal axes with the origin located 
at the center of site $j$.
The $E$2 amplitude is
\begin{eqnarray}
M_{2} (j:\textbf{k},\mbox{\boldmath{$\epsilon$}};
\textbf{k}',\mbox{\boldmath{$\epsilon$}}':\omega)
&=& \frac{k^2}{9} \sum_{\mu,\mu'=1}^{5}
q_{\mu}(\hat{\textbf{k}}' \cdot \mbox{\boldmath{$\epsilon$}}')
q_{\mu'}(\hat{\textbf{k}} \cdot \mbox{\boldmath{$\epsilon$}}) \nonumber \\
&\times& \sum_{\Lambda} 
\frac{\langle 0| \tilde{z}_{\mu,j} | \Lambda \rangle
\langle \Lambda | \tilde{z}_{\mu',j} | 0 \rangle}
{\hbar \omega - (E_{\Lambda}-E_0) + i \Gamma}, \label{eq.def.M2}
\end{eqnarray}
where factors $q_{\mu}(\hat{\textbf{k}}' \cdot \mbox{\boldmath{$\epsilon$}}')$ 
and $q_{\mu'}(\hat{\textbf{k}} \cdot \mbox{\boldmath{$\epsilon$}})$
with $\hat{\textbf{k}}'=\textbf{k}'/|\textbf{k}'|$ 
and $\hat{\textbf{k}}=\textbf{k}/|\textbf{k}|$ 
are defined as a second-rank tensor,
\begin{equation}
q_{\mu}(\textbf{A},\textbf{B})=\left\{
\begin{array}{ll}
\frac{\sqrt{3}}{2} (A_{x} B_{x} - A_{y} B_{y} ) & \textrm{for} \ \mu=1, \\
\frac{1}{2} (3A_{z} B_{z} - \textbf{A} \cdot \textbf{B}) & \textrm{for} \ \mu=2, \\
\frac{\sqrt{3}}{2} (A_{y} B_{z} + A_{z} B_{y} ) & \textrm{for} \ \mu=3, \\
\frac{\sqrt{3}}{2} (A_{z} B_{x} + A_{x} B_{z} ) & \textrm{for} \ \mu=4, \\
\frac{\sqrt{3}}{2} (A_{x} B_{y} + A_{y} B_{x} ) & \textrm{for} \ \mu=5. \\
\end{array} \right.
\end{equation}
Note that the quadrupole operator $\tilde{z}_{\mu,j}$ is expressed as
$\tilde{z}_{\mu,j}=q_{\mu}(\textbf{r}_{j},\textbf{r}_{j})$,
and the subscripts $\mu=1,2,3,4$, and $5$ for rank two quantity
specify the Cartesian components $x^2-y^2, 3z-r^2, yz, zx$, 
and $xy$, respectively.

In general, the evaluation of the scattering amplitude
$M_n$ tends to be a formidable task, since
the intermediate states of the
scattering process are difficult to calculate.
However, there exist several cases in which the evaluation
of the scattering amplitude becomes easy.
For instance, by replacing the energy denominators
in $M_{n}$ with a single oscillator, the amplitudes are
reduced into compact forms.\cite{Luo93,Lovesey05}
This treatment is called as '\textit{fast collision approximation}'.
Here, we adopt an another treatment, in which
the Hamiltonian describing the intermediate states is assumed to 
preserve a spherical symmetry. For only the $f$ states concerned,
this assumption is justified when the crystal electric field (CEF) 
and the intersite interaction are negligible compared with the intra-atomic 
Coulomb and the spin-orbit interactions in the intermediate states.
Analyses based on this framework gave good results in many localized 
$f$-electron systems.\cite{Nagao05,Nagao06,Nagao08}
In the present case, this assumption seems applicable to the intermediate 
state of the $E$2 transition, while the applicability is not clear for the 
intermediate states of the $E$1 transition, since the $5d$ bands are involved.
In this paper, assuming the same $5d$-DOS for the $e_g$ and $t_{2g}$ 
symmetries, we preserve the spherical symmetry in the intermediate states 
to calculate the RXS spectra.
This assumption is justified later in our semi-quantitative analysis,
since the RXS spectra are not sensitive to the shape and the fillingness
of the $5d$ DOS.

Within the present scheme, the RXS scattering amplitudes are expressed 
in neat forms suitable to discuss field dependence.
For the $E$1 process, the scattering amplitude at a single site is
\begin{equation}
M_{1} (\mbox{\boldmath{$\epsilon$}};\mbox{\boldmath{$\epsilon$}}':\omega)
= \sum_{\nu=0}^{2} \alpha_{1}^{(\nu)}(\omega)
\sum_{\mu=1}^{2 \nu+1} P_{1,\mu}^{(\nu)}
(\mbox{\boldmath{$\epsilon$}}, \mbox{\boldmath{$\epsilon$}}')
\langle 0| z_{\mu}^{(\nu)} | 0 \rangle, \label{eq.E1}
\end{equation}
where $z_{\mu}^{(\nu)}$ is operator equivalence of
mutipole moment of the
component $\mu$ with rank $\nu$.
For rank zero, $z_{1}^{(0)}=1$, and for rank one, $z_{\mu}^{(1)}$'s are
$J_{x}, J_{y}$, and $J_{z}$ with $\mu=1,2$, and $3$, respectively.
For rank two, quadrupole operator is represented by
$z_{\mu}^{(2)}=q_{\mu}(\textbf{J},\textbf{J})$.
The energy profile of rank $\nu$ contribution is
denoted as $\alpha_1^{(\nu)}(\omega)$, whose explicit form is
found in Ref. \onlinecite{Nagao05}.
The geometrical factors are given as follows:
for rank zero, $P_{1,1}^{(0)}
(\mbox{\boldmath{$\epsilon$}},\mbox{\boldmath{$\epsilon$}}')$$=$
$\mbox{\boldmath{$\epsilon$}}' \cdot\mbox{\boldmath{$\epsilon$}}$,
for rank one, $P_{1,\mu}^{(1)}
(\mbox{\boldmath{$\epsilon$}},\mbox{\boldmath{$\epsilon$}}')$$=$
$-i (\mbox{\boldmath{$\epsilon$}}' \times \mbox{\boldmath{$\epsilon$}} )_{\mu}$,
and for rank two, 
$P_{1,\mu}^{(2)}
(\mbox{\boldmath{$\epsilon$}},\mbox{\boldmath{$\epsilon$}}')$$=$
$q_{\mu}(\mbox{\boldmath{$\epsilon$}}', \mbox{\boldmath{$\epsilon$}} )$.
Note that we have omitted the subscript specifying the site
and do so hereafter.
For the $E$2 process, the scattering amplitude at a single site is
\begin{eqnarray}
M_{2} (\textbf{k},\mbox{\boldmath{$\epsilon$}};
\textbf{k}',\mbox{\boldmath{$\epsilon$}}':\omega)
&=&\frac{k^2}{9} 
\sum_{\nu=0}^{4} \alpha_{2}^{(\nu)}(\omega) 
\sum_{\mu=1}^{2 \nu+1}
P_{2,\mu}^{(\nu)}
(\mbox{\boldmath{$\epsilon$}},\mbox{\boldmath{$\epsilon$}}',
\hat{\textbf{k}},\hat{\textbf{k}}')
\nonumber \\
&\times& \langle 0 | z_{\mu}^{(\nu)} | 0 \rangle,
\label{eq.E2}
\end{eqnarray} 
where $P_{2,\mu}^{(\nu)}
(\mbox{\boldmath{$\epsilon$}},\mbox{\boldmath{$\epsilon$}}',
\hat{\textbf{k}},\hat{\textbf{k}}')$ is 
the geometrical factor of the
component $\mu$ with rank $\nu$.
The definitions of 
$z_{\mu}^{(\nu)}$ with rank $\nu$ higher than three 
and $P_{2,\mu}^{(\nu)}
(\mbox{\boldmath{$\epsilon$}},\mbox{\boldmath{$\epsilon$}}',
\hat{\textbf{k}},\hat{\textbf{k}}')$
are given in Ref. \onlinecite{Nagao06}.\cite{com.nagao}
These formulae look similar to those derived in 
literatures, mainly employing the fast collision
approximation.\cite{Hannon88,Luo93,Carra94,Hill96,Lovesey05}
Our treatment, however, is convenient when spectral analysis
is needed since the energy profiles in Eqs. (\ref{eq.E1}) 
and (\ref{eq.E2}) are correctly included.
In the next section, we shall see Eq. (\ref{eq.E2}) is strictly applicable 
to describing the $E$2 process at the Ce $L_{2,3}$ edges, 
while Eq. (\ref{eq.E1}) is approximately valid to express the $E$1 process.

\subsection{Intensity of the difference spectrum}
We investigate how the RXS intensity changes when
the orientation of the applied field is reversed.   
For antiferro-type scattering vector $\textbf{G}$,
the factor $\textrm{e}^{-i\textbf{G} \cdot \textbf{r}_j}$
appeared in Eq. (\ref{eq.amp.2}) gives $+1$ or $-1$ depending
on the kind of sublattice the site $j$ belonging to.
By substituting Eqs. (\ref{eq.amp.2}), (\ref{eq.E1}),
and (\ref{eq.E2}) into Eq. (\ref{eq.amp.1}),
we obtain
the total amplitude of RXS with $\textbf{H}$ along a certain 
direction as
\begin{eqnarray}
& & F(\textbf{k},\mbox{\boldmath{$\epsilon$}};
\textbf{k}',\mbox{\boldmath{$\epsilon$}}':\omega)
\propto \sqrt{N} \left[
i \alpha_{1}^{(1)}(\omega) Z_{1}^{(1)} +
\alpha_{1}^{(2)}(\omega) Z_{1}^{(2)} \right. \nonumber \\
& & \left. + i \alpha_{2}^{(1)}(\omega) Z_{2}^{(1)} +
\alpha_{2}^{(2)}(\omega) Z_{2}^{(2)}
+ i \alpha_{2}^{(3)}(\omega) Z_{2}^{(3)} \right],
\end{eqnarray}
where the term $\alpha_{2}^{(4)}(\omega) Z_{2}^{(4)}$ is omitted
since it is absorbed into $\alpha_{2}^{(2)}(\omega) Z_{2}^{(2)}$
in the present case of CeB$_{6}$.\cite{Shiina97}
Here, use has been made of a new quantity
\begin{equation}
Z_{n}^{(\nu)} = \left\{ \begin{array}{rl}
\sum_{\mu=1}^{2 \nu+1} P_{n,\mu}^{(\nu)}
(\mbox{\boldmath{$\epsilon$}},\mbox{\boldmath{$\epsilon$}}')
\langle z_{\mu}^{(\nu)} \rangle 
& \textrm{for} \ \ \nu=\textrm{even} \\
-i \sum_{\mu=1}^{2 \nu+1} P_{n,\mu}^{(\nu)}
(\mbox{\boldmath{$\epsilon$}},\mbox{\boldmath{$\epsilon$}}')
\langle z_{\mu}^{(\nu)} \rangle 
& \textrm{for} \ \ \nu=\textrm{odd} \\
\end{array} \right. ,
\end{equation}
where  the staggered moment is referred to as 
$\langle z_{\mu}^{(\nu)} \rangle$. It is related to the sublattice
moments as $\langle z_{\mu}^{(\nu)} \rangle$$=$
$\langle 0| z_{\mu}^{(\nu)}|0 \rangle_{A}$$=$
$-\langle 0| z_{\mu}^{(\nu)}|0 \rangle_{B}$ where
the subscripts $A$ and $B$ distinguish the sublattices.
We emphasize that $\{ Z_{n}^{(\nu)} \}$'s are real quantities.
Note that only staggered components of the multipole operators
contribute to the amplitude with the antiferro-type $\textbf{G}$.

The RXS intensity $I(\omega,\textbf{H})$ is given by 
$|F(\textbf{k},\mbox{\boldmath{$\epsilon$}};
\textbf{k}',\mbox{\boldmath{$\epsilon$}}':\omega)|^{2}$.
When the direction of the applied field is reversed, 
$Z_{1}^{(\nu)}$ and $Z_{2}^{(\nu)}$ having odd $\nu$ reverse 
their signs, while those having even $\nu$ remain unchanged.
Therefore, the amplitude for $-\textbf{H}$ is expressed 
by the quantities for $+\textbf{H}$. Then,
the total intensities are expressed as
\begin{eqnarray}
& & I(\omega, \pm \textbf{H})
\propto \left|\alpha_{1}^{(2)}(\omega) Z_{1}^{(2)}
+ \alpha_{2}^{(2)}(\omega) Z_{2}^{(2)} \right. \nonumber \\
& & \left.
\pm i \{ \alpha_{1}^{(1)}(\omega) Z_{1}^{(1)} +
\alpha_{2}^{(1)}(\omega) Z_{2}^{(1)} +
\alpha_{2}^{(3)}(\omega) Z_{2}^{(3)} \} \right|^2. \label{eq.intensity.1}
\end{eqnarray}
The difference spectrum is
defined as
\begin{equation}
\Delta I(\omega) \equiv 
\frac{I(\omega, +\textbf{H}) -I(\omega, -\textbf{H})}{2}.
\label{eq.diff.spec}
\end{equation}
The spectrum $\Delta I(\omega)$ produces non-zero contribution
when cross terms between the amplitude with odd rank and that
with even rank remains finite.
By substituting (\ref{eq.intensity.1}) into Eq. (\ref{eq.diff.spec}), 
we can classify such cross terms into three categories:
\begin{eqnarray}
\Delta I (\omega) &=& \Delta I_{E1E2} + \Delta I_{E1E1} + \Delta I_{E2E2}, \\
\Delta I_{E1E2} &\equiv&
    2 Z_{1}^{(1)} Z_{2}^{(2)} 
\textrm{Im} [ \{ \alpha_{1}^{(1)}(\omega) \}^{\star}
                 \alpha_{2}^{(2)}(\omega) ] \nonumber \\
&+& 2 Z_{1}^{(2)} Z_{2}^{(1)} 
\textrm{Im} [ \{ \alpha_{2}^{(1)}(\omega) \}^{\star}
                 \alpha_{1}^{(2)}(\omega) ] \nonumber \\
&+& 2 Z_{1}^{(2)} Z_{2}^{(3)} 
\textrm{Im} [ \{ \alpha_{2}^{(3)}(\omega) \}^{\star}
                 \alpha_{1}^{(2)}(\omega) ], \label{eq.E1E2} \\
\Delta I_{E1E1}
&\equiv& 2 Z_{1}^{(1)} Z_{1}^{(2)} 
\textrm{Im} [ \{ \alpha_{1}^{(1)}(\omega) \}^{\star}
                 \alpha_{1}^{(2)}(\omega) ], \label{eq.E1E1} \\
\Delta I_{E2E2} &\equiv&
2 Z_{2}^{(1)} Z_{2}^{(2)} 
\textrm{Im} [ \{ \alpha_{2}^{(1)}(\omega) \}^{\star}
                 \alpha_{2}^{(2)}(\omega) ] \nonumber \\
&+& 2 Z_{2}^{(3)} Z_{2}^{(2)} 
\textrm{Im} [ \{ \alpha_{2}^{(3)}(\omega) \}^{\star}
                 \alpha_{2}^{(2)}(\omega) ], \label{eq.E2E2}
\end{eqnarray}
where $\textrm{Im}[X]$ stands for imaginary part of $X$.

Before showing numerical results of the RXS spectra,
we comment on an outcome expected from the fast collision 
approximation, in which the energy denominators
in Eqs. (\ref{eq.def.M1}) and (\ref{eq.def.M2}) are
factored out of the summation over the intermediate states 
and replaced by a single oscillator.\cite{Luo93,Lovesey05}
As a consequence, the energy profile $\alpha_{n}^{(\nu)}(\omega)$
loses its $\nu$ dependence, say $\alpha_{n}(\omega)$.
Then, $\Delta I_{E1E1}(\omega)$ and $\Delta I_{E2E2}(\omega)$
become zero, and $\Delta I_{E1E2}(\omega)$ alone
remains as
\begin{eqnarray}
\Delta I_{E1E2}(\omega) &=&
2 [ Z_{1}^{(1)} Z_{2}^{(2)} - Z_{1}^{(2)}( Z_{2}^{(1)} + Z_{2}^{(3)})]
\nonumber \\
& & \times
 \textrm{Im} [ \{ \alpha_{1}(\omega) \}^{\star} \alpha_{2}(\omega) ].
\end{eqnarray}
This expression is what Matsumura \textit{et al}. have used in their 
analysis.\cite{Matsumura09}

\section{Initial and intermediate states \label{sect.3}}
Cerium hexaboride is believed to show an AFQ ordering phase
in the temperature range $T_{\textbf{N}} \leq T \leq T_{\textrm{Q}}$
with $T_{\textrm{N}}=2.3$ K\cite{Effantin85,Zaharko03} and
$T_{\textrm{Q}}=3.3$ K under no applied magnetic 
field.\cite{Komatsubara80,Effantin85,Yakhou01}
It shows a simple cubic structure (CsCl-type,
P$_{m\overline{3}m}$) with 
lattice constant $a$ being $4.14 \textrm{\AA}$.
In CeB$_{6}$, Ce is trivalent in the $f^{1}$ configuration 
forming a sextet term $^{2}\textrm{F}_{\frac{5}{2}}$.
Under the cubic symmetry CEF, the sextet splits into a $\Gamma_{7}$
doublet and a $\Gamma_{8}$ quartet. 
The latter is the lowest energy state.
Since the energy splitting between them
is on the order of 530 K,\cite{Zirngiebl84}
the $\Gamma_{8}$ quartet alone is sufficient for
investigating low-temperature phenomena.
Within the $\Gamma_{8}$ basis under the $O_{h}$ symmetry, 
multipolar operators with rank one, two, and three are active.
In the following, we introduce a model Hamiltonian defined
in a subspace spanned by $\Gamma_{8}$ basis.

\subsection{Model Hamiltonian}
In order to prepare the initial (or ground) state of the AFQ state
in CeB$_6$, we adopt the model Hamiltonian utilized in our previous
works.\cite{Nagao01,Igarashi02,Nagao03}
It is originally introduced by Ohkawa\cite{Ohkawa85} and
extended by Shiina \textit{et al}.\cite{Shiina97}
It is derived on the basis of the RKKY interaction
and a possibility of the anisotropic RKKY interaction is discarded for 
simplicity.\cite{Schlottmann00}
The Hamiltonian is 
\begin{eqnarray}
\hat{H}&=& \frac{D}{4} \sum_{\langle i,j \rangle}
\left[ ( 1+\delta )\sum_{\mu=3}^{5} O_{\mu,i} O_{\mu, j}
+ \frac{1}{16} \sum_{\mu=1}^{2}O_{\mu, i} O_{\mu, j} \right] \nonumber \\
&+& D \sum_{\langle i,j \rangle}
\left[ \tau_{i}^{y} \tau_{j}^{y}
+ \mbox{\boldmath{$\sigma$}}_{i} \cdot \mbox{\boldmath{$\sigma$}}_{j}
+ \mbox{\boldmath{$\eta$}}_{i} \cdot \mbox{\boldmath{$\eta$}}_{j}
+ \mbox{\boldmath{$\zeta$}}_{i} \cdot \mbox{\boldmath{$\zeta$}}_{j} \right] 
\nonumber \\
&+& g \mu_{\textrm{B}}\sum_{i} \textbf{J}_{i} \cdot \textbf{H},
\label{eq.Hamiltonian}
\end{eqnarray}
where operator equivalence of quadrupole moment is defined as
$O_{\mu}= q_{\mu}(\textbf{J},\textbf{J})$.
The second line of Eq. (\ref{eq.Hamiltonian}) describes the
interactions between dipole and octupole moments,
and the definitions of the symbols appeared in this line
are found, e.g., in Refs. \onlinecite{Shiina97} and \onlinecite{Igarashi02}.
The sum on $\langle i,j \rangle$ runs over nearest neighbor Ce pairs.
The last line in Eq. (\ref{eq.Hamiltonian}) stands for the Zeeman term
with $g$-factor being $6/7$.
Note that, by choosing parameter $\delta$ positive,
this Hamiltonian favors the AFQ order of three components belonging to
the $\Gamma_{5}$ basis ($O_{yz}, O_{zx}$, and $O_{xy}$).
The parameter $\delta$ is fixed as $\delta=0.2$.
The coupling constant is chosen so as to reproduce the
value of $T_{\textrm{Q}}$ in the absence of magnetic field.

\subsection{Mean field solutions and the initial states}
\begin{figure}[t]
\begin{center}\includegraphics[width=8.0cm]{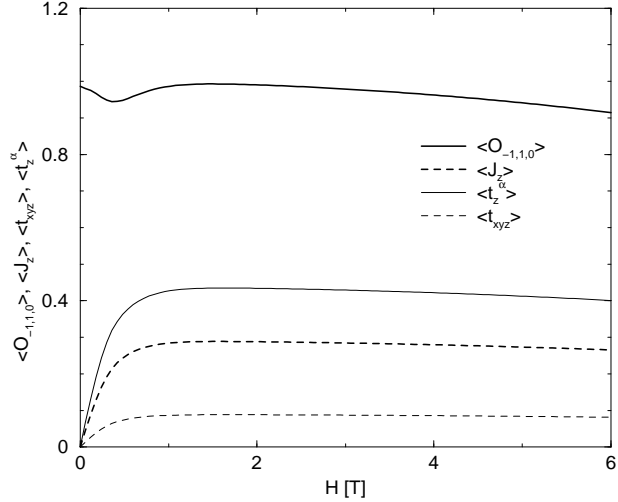}
\caption{H-dependence of the staggered moments at $T=1.65$ K.
Bold solid and broken lines are $\langle O_{-1,1,0} \rangle$
and $\langle J_z \rangle$, respectively.
Thin solid and broken lines denote
$\langle t_{z}^{\alpha} \rangle=\langle T_{z}^{\alpha} \rangle/\sqrt{50}$
and $\langle t_{xyz} \rangle=\langle T_{xyz} \rangle/(45/2\sqrt{5})$.
\label{fig.af}}
\end{center}
\end{figure}

We apply the mean field approximation to the
Hamiltonian.
Under the influence of the external field $\textbf{H}$ along $(h,k,\ell)$
direction, the mean field solution gives a ground state of the primary
order parameter 
$\langle O_{hk\ell}\rangle = 
\langle (h O_{yz} + k  O_{zx} + \ell O_{xy})\rangle/\sqrt{h^2+k^2 +\ell^2}$.
The quadrupole ordering temperature $T_{\textrm{Q}}^0$ in the zero field limit
is given by 3.3 K for $D=0.458$ K.
The field induces another ranks and/or another components
of multipole order parameters depending on the direction
of the applied field.\cite{Shiina97,Shiina98}
In the case of $\textbf{H} \parallel (\overline{1},1,0)$ adopted
by Matsumura \textit{et al}.,\cite{Matsumura09}
three antiferro-type of the secondary order parameters, 
a dipole component $J_z$ and octupole components $T_{xyz}$ and
$T_{z}^{\alpha}$, are induced. Note that
the definitions of $T_{xyz} (=z_{1}^{(3)})$ and
$T_{z}^{\alpha} (=z_{4}^{3)}$) are given in Ref. \onlinecite{Nagao06}.
The field dependences of these order parameters
are shown in Fig. \ref{fig.af} for $T=T_{\textrm{Q}}^0/2=1.65$ K.
Since $|\textbf{H}|$-dependence of the order parameters, 
both primary and induced ones, become mild beyond
$|\textbf{H}| \simeq 1$T, we will fix the field
strength $|\textbf{H}|=2$ T throughout the present work. 
This choice of the magnitude of $\textbf{H}$ 
enables us to avoid complication concerning
the problem of domain population, which is known to exist when
the applied field is much smaller as seen in 
Fig. 2 of Ref. \onlinecite{Matsumura09}.
Finally, note that though many more order parameters are induced 
in ferro-type alignments,
we do not mention them since they have no contribution
on the RXS intensity at the antiferro-type scattering vector
now addressed.

\subsection{Intermediate states}
\begin{figure}[t]
\begin{center}\includegraphics[width=8.0cm]{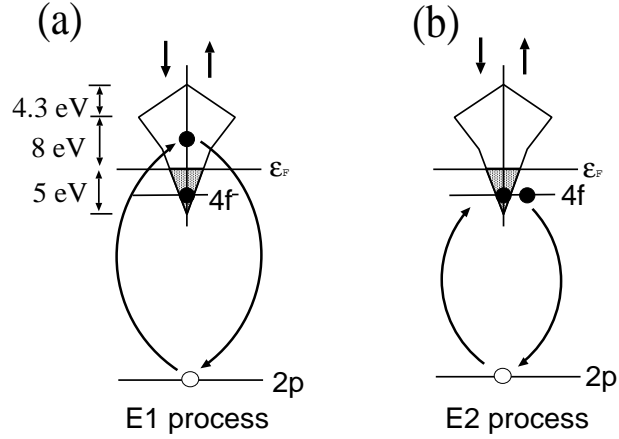}
\caption{Schematic diagrams of two RXS processes at the Ce
$L_{2}$ and $L_{3}$ edges:
(a) $E$1 process; (b) $E$2 process
The $5d$ DOS is schematically exhibited in the upper part of the figure
and arrows indicate spin of the $5d$ states.
\label{fig.intermediate}}
\end{center}
\end{figure}

The intermediate states of RXS near the Ce $L_{2,3}$ absorption edges
include excitations of an electron from core $2p$ states at a Ce site 
into the $5d$ and $4f$ states in the $E$1 and $E$2 processes,
respectively.
Since the $5d$ states form not levels but conduction bands,
we need a model of the density of states (DOS) of them.
In order to keep continuity from our previous work,
we employ the same $5d$ DOS used before.\cite{Nagao01,Igarashi02,Nagao03}
The $5d$ DOS, $\rho^{5d}(x)$, is assumed to be
\begin{equation}
\rho^{5d}(x) = \left\{ \begin{array}{ll}
0.008  x + 0.04, & -5 < x < 0, \\
0.01 x + 0.04, & 0 < x < 8 \\
-0.0277 x + 0.342, & 8 < x < 12.33,
\end{array} \right.
\end{equation}
where $x$ is measured in units of eV with $x=0$ corresponding to
the Fermi level.  Total number of the occupied $5d$ electron per
Ce site is set to be unity.  
We disregard the dependence on the $5d$ states
$x^2-y^2, 3z^2-r^2, yz,zy$, and $xy$.
These settings are justified later in our semi-quantitative analysis,
since the RXS spectra are not sensitive to the shape and the fillingness
of the $5d$ DOS.
Figure \ref{fig.intermediate} shows a schematic view
of the RXS processes and the shape of the $5d$ DOS. 

\subsubsection{$E$1 process}
As explained above, the $E$1 transition at the $L_{2,3}$ edges 
consists of that between the $2p$ and $5d$ states.
We consider the resolvent $[\hbar \omega - H_{int}]^{-1}$,
where $H_{int}$ is the Hamiltonian spanned in the configuration
involving one $4f$ electron, one $2p$ core hole, and one
excited electron in the $5d$ band.
First, we solve an eigenvalue problem
considering the Coulomb interaction between the $4f$ and $2p$ core hole 
as well as the spin-orbit interaction of them at the central site.
Let the eigenvalue and the eigenstate be $E_{\lambda}$
and $| \lambda \rangle$, respectively.
In this work, 
the Slater integrals and the spin-orbit interaction
parameters needed to evaluate the Coulomb and spin-orbit
interactions are calculated for Ce$^{3+}$ atom 
within the Hartree-Fock (HF) approximation.\cite{Cowan81}
The obtained off-diagonal and diagonal values of the Slater integrals
are multiplied by factors 0.8 and 0.25, respectively,
taking the screening effect into account.
Then, the presence of the $5d$ electron is treated as a scattering problem.
The problem is described by an inverse matrix problem
symbolically summarized below:
\begin{eqnarray}
& & \left(\frac{1}{\hbar \omega - H_{int} + i \gamma}
 \right)_{d \lambda,d' \lambda'} \nonumber \\
&=& \left[ \left\{ G^{5d}(\hbar \omega + i \Gamma - E_{\lambda})\right\}^{-1} 
\delta_{d \lambda,d' \lambda'}
- V_{d \lambda,d' \lambda'} \right]^{-1},
\end{eqnarray}
where $d=(m_d,s_d)$ specifies a state of the $5d$ electron.
Matrix $V$ stands for the Coulomb interactions between the $5d$ and $4f$
electrons and between the $5d$ electron and the $2p$ core hole.
The local Green function of the $5d$ electron $G~{5d}(\omega)$ is 
defined by
\begin{equation}
G^{5d}(\hbar \omega) = \int_{0}^{\infty} \frac{\rho^{5d}(\epsilon)}
{\hbar \omega - \epsilon + i \gamma} d \epsilon,
\gamma \to 0.
\end{equation}
The RXS amplitude is calculated with rewriting Eq.~(\ref{eq.def.M1}) as
\begin{equation}
M_{1} (j:\mbox{\boldmath{$\epsilon$}};\mbox{\boldmath{$\epsilon$}}':\omega)
=\sum_{\mu,\mu'=1}^{3} \epsilon_{\mu}' \epsilon_{\mu'}
\sum_{d\lambda,d'\lambda'}
\langle 0| x_{\mu,j} |d\lambda \rangle
\left(\frac{1}{\hbar\omega-H_{int}}\right)_{d\lambda,d'\lambda'}
\langle d'\lambda' | x_{\mu',j} | 0 \rangle. \label{eq.3.5}
\end{equation}
Detail of the resolvent treatment is found in Ref. \onlinecite{Igarashi02}.
Although the original derivation of Eq. (\ref{eq.E1}) 
in Ref. \onlinecite{Nagao05} does not
expect inclusion of the $5d$ band, the form of Eq. (\ref{eq.E1}) is still 
correct, since the $5d$ DOS possesses a spherical nature. 
The dipole matrix element $A_{dp}$$=$$\langle 5d | r|2p \rangle$
$=$$\int_{0}^{\infty}R_{\textrm{5d}}(r)r R_{\textrm{2p}}(r) r^{2} dr$
is included implicitly in Eq. (\ref{eq.3.5}) where
$R_{\textrm{2p}}(r)$ and $R_{\textrm{5d}}(r)$ are the radial wave functions
for the $2p$ and $5d$ states, respectively.
Within the HF approximation, it is evaluated as
$A_{dp}=3.67 \times 10^{-11} \textrm{cm}$.\cite{Cowan81}

\subsubsection{$E$2 process}
The intermediate states in the $E$2 process can be constructed within the
$(2p)^5 (4f)^2$ configuration, disregarding the $5d$ electrons preoccupied
in the ground state.
The Hamiltonian describing the intermediate states
consists of the intra-atomic Coulomb and spin-orbit interactions
in this configuration.
The Slater integrals and the spin-orbit interaction
parameters are calculated within the HF
approximation.\cite{Cowan81}
The obtained off-diagonal and diagonal values of the Slater integrals
are multiplied by factors 0.8 and 0.25, respectively,
taking the screening effect into account.
The Hamiltonian matrix is numerically diagonalized by representing it
in the $(2p)^5 (4f)^2$ configuration. The RXS amplitude is calculated by
inserting the eigenstates and eigenvalues into Eq.~(\ref{eq.def.M2}).
Note that the scattering amplitude of this process is written 
by Eq. (\ref{eq.E2}) since the Hamiltonian preserves the spherical symmetry.
In Eq. (\ref{eq.E2}), the quadrupole matrix element 
$A_{fp}$$=$$\langle 4f | r^{2}|2p \rangle$
$=$$\int_{0}^{\infty}R_{\textrm{4f}}(r)r^{2} R_{\textrm{2p}}(r) r^{2} dr$
is implicitly included where
$R_{\textrm{4f}}(r)$ denotes the radial wave function for the $4f$ state.
Within the HF approximation, it is evaluated as
$A_{fp}=5.69 \times 10^{-20} \textrm{cm}^{2}$.\cite{Cowan81}

\section{Numerical results \label{sect.4}}

\begin{figure}[t]
\begin{center}\includegraphics[width=8.0cm]{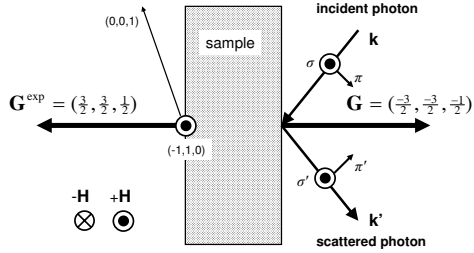}
\caption{Schematic diagrams of the RXS geometry.
The setting is adopted from Ref. \onlinecite{Matsumura09}
\label{fig.geometry}}
\end{center}
\end{figure}

In the following, we shall report the numerical results of the
RXS spectra.  First, we clarify the setting of our RXS calculation.
A schematic aspect is found in Fig. \ref{fig.geometry}, in which 
the scattering vector $\textbf{G}$
and the photon polarization are depicted.
Contrary to our definition of $\textbf{G}$, some literatures,
including the experimental works we analyze 
in the following,\cite{Matsumura09,Nakao01}
adopt the opposite sign, i.e., 
$\textbf{G}^{\textrm{exp}}=\textbf{k}-\textbf{k}'$.
To avoid a confusion,
when we mention $\textbf{G}$ in this section, 
we mean $\textbf{k}-\textbf{k}'$, while the
actual calculations are carried out by using $\textbf{k}'-\textbf{k}$,
because the formulae derived in our previous papers
are the results of the latter definition.

\subsection{At $\textbf{G}=(\frac{3}{2}\frac{3}{2}\frac{1}{2})$ 
under $\textbf{H}//(\overline{1}10)$}
\subsubsection{at the Ce $L_{3}$-edge}
\begin{figure}[t]
\begin{center}\includegraphics[width=8.0cm]{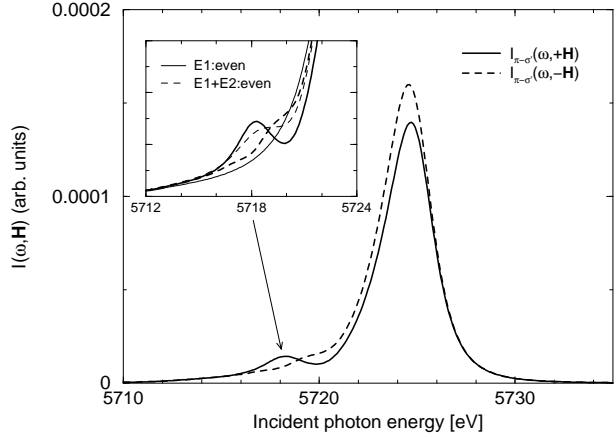}
\caption{RXS spectra around the Ce $L_{3}$ edges
at $\textbf{G}=(\frac{3}{2}\frac{3}{2}\frac{1}{2})$
with $\textbf{H} \parallel (\overline{1}10)$.
Bold solid and bold broken lines represent 
$I_{\pi-\sigma'}(\omega,+\textbf{H})$ and $I_{\pi-\sigma'}(\omega,-\textbf{H})$,
respectively.
Inset: $I(\omega,\pm \textbf{H})$ around the $E$2 peak. 
Thin solid and thin broken lines represent 
$|\alpha_{1}^{(2)}(\omega)|^{2} [Z_{1}^{(2)}]^2$
and $|\alpha_{1}^{(2)}(\omega) Z_{1}^{(2)}
+ \alpha_{2}^{(2)}(\omega) Z_{2}^{(2)}|^{2}$,
respectively.
\label{fig.L3}}
\end{center}
\end{figure}

Matsumura \textit{et al}. carried out the RXS experiment under the
applied magnetic field $\textbf{H}$ along $(\overline{1}10)$ direction near 
Ce $L_3$ absorption edge at $\textbf{G}=(\frac{3}{2}\frac{3}{2}\frac{1}{2})$
in the $\pi$-incident polarization.\cite{Matsumura09}
They found the spectra showed strong enhancement of the intensity
around the $E$1 and $E$2 regions.
They also observed that the spectral shape underwent the significant 
change when the orientation of $\textbf{H}$ was reversed.
In particular, due to 
the cross term between the even rank and odd rank contributions,
the spectra showed two-peak structure with peaks at the $E$1 and $E$2
positions when $\textbf{H}$ was along $(\overline{1}10)$, while
they showed single-peak structure with the $E$2 peak merged into 
the tail part of the $E$1 signal when $\textbf{H}$ was along $(1\overline{1}0)$.
We calculated the RXS spectra with the settings adjusted to those
of Matsumura \textit{et al}'s (Fig. \ref{fig.geometry}).  
The results are shown in Fig. \ref{fig.L3}.  
The origins of the energy are set and fixed such that
the $E$1 and $E$2 peaks become around 5724 eV and 5718 eV, respectively.

First, we concern the whole aspect of the spectral shapes.
The calculated curves capture the above explained experimental feature well.
We also confirm that the similar tendency is expected from the 
spectra in the $\sigma-\pi'$ channel (not shown) with the intensity
of the $E$1 peak being roughly a fourth of that in the $\pi-\sigma'$ channel.
Note that, the ratio of the intensity at the $E1$ peak to that at the $E$2
peak seems apparently quite different between the experiment and our result.
For instance, for $+\textbf{H}$ ($|\textbf{H}|=2$ T),
the ratio is about 2.45 in the experiment\cite{Matsumura09} 
and about 9.66 in our calculation.
A remedy for this discrepancy is absorption correction.
The experimental ratio is enhanced to about 10.5
after the correction is properly carried out.\cite{com.Matsumura.2}
So, our theoretical ratio gives fairly good values.

Next, we analyze the ingredients of the intensity around 
the $E$1 and $E$2 positions. To this aim, we define average intensity as
\begin{equation}
I_{\textrm{av}}(\omega)=\frac{I(\omega,+\textbf{H})+I(\omega,-\textbf{H})}{2}.
\end{equation}
At the $E$1 peak, by using Eq. (\ref{eq.intensity.1}),
the average intensity is approximated as
\begin{equation}
I_{\textrm{av}}(\omega) \simeq |\alpha_{1}^{(2)}(\omega)|^2 [Z_{1}^{(2)}]^2,
\end{equation}
since another $E$1 contribution of $|\alpha_{1}^{(1)}(\omega)|^2 [Z_{1}^{(1)}]^2$
is two orders of magnitude smaller than that of
$|\alpha_{1}^{(2)}(\omega)|^2 [Z_{1}^{(2)}]^2$. 
At the $E$2 peak, the situation is not so simple.
As shown in the inset of Fig. \ref{fig.L3}, about half of the
intensity around the $E$2 position is supplied by
the tail part of the $E$1 contribution, 
$|\alpha_{1}^{(2)}(\omega)|^2 [Z_{1}^{(2)}]^2$.
Other than that, an even-rank profile of the $E$2 transition
$\alpha_{2}^{(2)}(\omega)$
also has contribution by interfering with
$\alpha_{1}^{(2)}(\omega)$ as illustrated by thin broken line.
That is,
\begin{equation}
I_{\textrm{av}}(\omega) \simeq | \alpha_{1}^{(2)}(\omega) Z_{1}^{(2)}
+ \alpha_{2}^{(2)}(\omega) Z_{2}^{(2)}|^2.
\end{equation}

\begin{figure}[t]
\begin{center}\includegraphics[width=8.0cm]{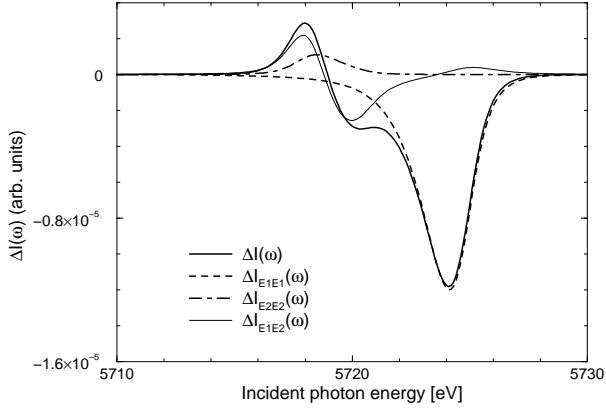}
\caption{Difference spectra around the Ce $L_{3}$ edges
at $\textbf{G}=(\frac{3}{2}\frac{3}{2}\frac{1}{2})$
with $\textbf{H} \parallel (\overline{1}10)$ in the $\pi-\sigma'$ channel.
Bold solid line represents the total $\Delta I(\omega)$.
Bold broken, bold dot-dashed, and thin solid lines are
$\Delta I_{E1E1}(\omega), \Delta I_{E2E2}(\omega)$, and $\Delta I_{E1E2}(\omega)$,
respectively.
\label{fig.diff}}
\end{center}
\end{figure}

Finally, we turn our attention to the difference 
spectrum $\Delta I(\omega)$ displayed in Fig. \ref{fig.diff}. 
The sign around $E$1 peak and that around $E$2 peak 
are opposite to each other, which is also in accordance with
the experiment.\cite{Matsumura09} 
Note that, as mentioned in Sec. \ref{sect.1},
although the $\Delta I(\omega)$ around the $E$1 region
is not so prominent in Ref. \onlinecite{Matsumura09},
its existence is assured by more careful measurement.\cite{com.Matsumura}
The presence of $\Delta I(\omega)$ intensity at the $E$1 position, in itself,
is ascribed to Eq. (\ref{eq.E1E1}) of $\Delta I_{E1E1}$ $\propto 
\langle J_z \rangle \langle O_{\overline{1},1,0} \rangle$. 
It consists of the cross term between rank one and rank two contributions
arose both from the $E$1 transition, which is missing within the
fast collision approximation as stated before.

On the other hand, the origin of the intensity around the $E$2 peak
is rather complicated.
The whole shape is determined by $\Delta I_{E1E2}$, while
both $\Delta I_{E1E1}$ and $\Delta I_{E2E2}$
have several quantitative contributions too.
In the present setting, $Z_{2}^{(3)}$ is two orders of magnitude
larger than $Z_{2}^{(1)}$ and the former is predominated by the contribution of
$T_{xyz}$. Then, $\Delta I_{E1E2}(\omega)$ and
$\Delta I_{E2E2}(\omega)$ are approximated as
$2 Z_{1}^{(2)} Z_{2}^{(3)} \textrm{Im} 
[ \{ \alpha_{2}^{(3)}(\omega) \}^{\star} \alpha_{1}^{(2)}(\omega) ]$ 
and $2 Z_{2}^{(3)} Z_{2}^{(2)} \textrm{Im} 
[ \{ \alpha_{2}^{(3)}(\omega) \}^{\star}\alpha_{2}^{(2)}(\omega) ]$, 
respectively.  
Therefore, the entire spectral shape is well controlled by
three energy profiles $\alpha_{1}^{(2)}(\omega)$, $\alpha_{2}^{(2)}(\omega)$, 
and $\alpha_{2}^{(3)}(\omega)$ carrying by the 
AFQ component of $O_{\overline{1},1,0}$ and AFO component of $T_{xyz}$.

Here, we end this subsection with the explanations concerning the
robustness of the present results.
Among several assumptions we have employed in the present
work, the choice of the $5d$ DOS seems the most crucial one.
The reason we have adopted the same DOS as used in our previous 
works is partly because to preserve the spherical symmetry in the intermediate
states and partly to keep continuity of our research.
Furthermore, we have confirmed that the characteristic features
our data have shown are insensitive to the modification of the shape of
the $5d$ DOS, fillingness of the $5d$ electron, and
presence or absence of the DOS splitting between the
$e_{g}$ and $t_{2g}$ states. 
For example, we have examined the semi-elliptic shape DOS
and the uniform DOS. We also have changed the fillingness
from 0.5 to 1.8 per Ce site and have introduced the DOS splitting
between the $e_{g}$ and $t_{2g}$ states from $-3$ to $+3$ eV
keeping the shape of the $5d$ DOS the same as that in Fig. \ref{fig.intermediate}.
Although these modifications cause minor quantitative
differences, the main features we stated in this section 
remain unchanged.

\subsubsection{at the Ce $L_{2}$-edge}
\begin{figure}[h]
\begin{center}\includegraphics[width=8.0cm]{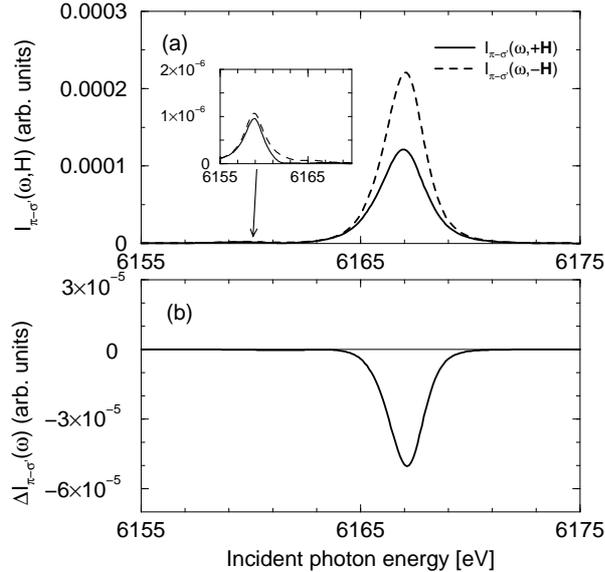}
\caption{RXS spectra around the Ce $L_{2}$ edges
at $\textbf{G}=(\frac{3}{2}\frac{3}{2}\frac{1}{2})$
with $\textbf{H} \parallel (\overline{1}10)$, respectively.
(a) Solid and broken lines represent $I_{\pi\sigma'}(\omega,+\textbf{H})$
and $I_{\pi\sigma'}(\omega,-\textbf{H})$
in the $\pi-\sigma'$ channels.
Inset:  $I(\omega,\pm \textbf{H})$ around the $E$2 peak.
(b) $\Delta I_{\pi\sigma'}(\omega)$ 
in the $\pi-\sigma'$ channel.
\label{fig.L2}}
\end{center}
\end{figure}

Next, we also calculate
the RXS spectra and the difference spectra
expected from the AFQ phase in the vicinity of the Ce $L_2$ edge
where the $E$1 and $E2$ transitions are observed around
6167 eV and 6160 eV, respectively. The results are shown in
Fig. \ref{fig.L2}.
The intensities are nearly the same as or slightly
stronger than those at the $L_3$ edge (Fig. \ref{fig.L3}).
We conclude that the RXS signal is experimentally detectable
near the Ce $L_2$ absorption edge.
Then, we concern the spectral shapes.
One notable feature is the $E$2 process has no practical contribution.
This shows a striking difference compared to the RXS spectra detected
near the Ce $L_2$ edge
in the antiferro-octupolar (AFO) phase of Ce$_{1-x}$La$_{x}$B$_6$
where the signals from the $E$2 transition were distinctly
observed.\cite{Mannix05,Kusunose05,Nagao06,Lovesey07}
The difference is attributed to that of the primary order parameter
in these systems.

\subsection{At $\textbf{G}=(\frac{1}{2}\frac{1}{2}\frac{1}{2})$ 
under $\textbf{H}//(11\overline{2})$}

Nakao \textit{et al}. detected the RXS signal of Ce $L_{3}$ edge 
from the AFQ phase.\cite{Nakao01}
However, their result practically showed no $E$2 contribution contrary to the
present case reported by Matsumura \textit{et al}., in which the peak intensity
at the $E$2 transition has distinguishable contribution 
from that at the $E$1 transition.\cite{Matsumura09}
It is reasonable to interpret the difference is due 
to that of experimental conditions
since the experiment by Nakao \textit{et al}. was 
carried out at $\textbf{G}=(\frac{1}{2}\frac{1}{2}\frac{1}{2})$ 
under $\textbf{H}//(11\overline{2})$ and the polarization of the incident
photon was $\sigma$ channel, all of which are different from the conditions
chosen by Matsumura \textit{et al}. 

In our previous works, we analyzed Nakao \textit{et al}.'s data and
confirmed that the peak intensity at the
$E$2 transition was negligible compared 
with the one at the $E$1 transition.\cite{Nagao01,Igarashi02}
In these treatments, we set the peak position of $E$2 transition
about ten eV lower than that of $E$1 transition.
It turns out that
the interval is too wide, since it is about six eV as seen 
from Matsumura \textit{et al}.'s data.
If we set the $E$2 peak position at six eV lower than
that of $E$1 transition, both intensities may experience interference
even under the Nakao \textit{et al}.'s experimental conditions.
However, the calculated results (not shown), practically, 
have neither notable peak nor $\Delta I(\omega)$ 
intensities around the $E$2 position, which confirm our previous
results survive after the shift of the $E$2 position.
The absence of the distinct peak at the $E$2 position
is merely the numerical reason.
The tail part of the $E$1 contribution around the $E$2
position in Nakao \textit{et al}.'s case is nearly twice larger than
that in Matsumura \textit{et al}.'s case.
The former buries the $E$2 contribution, which is nearly
the same in the latter case.

\section{Concluding Remarks}

We have theoretically investigated 
the RXS spectra observed in the AFQ phase of CeB$_{6}$ in the 
vicinity of the Ce $L_{3}$ edge at 
$\textbf{G}=\left(\frac{3}{2},\frac{3}{2},\frac{1}{2}\right)$ under the
applied field $\textbf{H}$ along $\parallel (\overline{1},1,0)$ direction.
The experimental data show small but clear contribution from the $E$2 process
as well as that from the main $E$1 process.
Also the interference between rank even and rank odd contributions
from the $E$1 and/ or $E$2 signals are observed,
which provides a great opportunity to obtain the information of the 
field-induced multipole orderings .
To analyze the RXS spectra, we have employed the model
on the basis of a localized electron picture, which 
is used in explaining several aspects of the RXS phenomena in CeB$_{6}$ and 
Ce$_{1-x}$La$_{x}$B$_{6}$.\cite{Nagao01,Igarashi02,Nagao03,Nagao06} 
The model is combined with the intermediate states including the intra-atomic
Coulomb and the spin-orbit interactions within the appropriate
electron configurations, so that the RXS signal is brought about by
the Coulomb interaction. 

The calculated spectra successfully 
capture the characteristic
features the experimental data show:
the ratio between the $E$1 and $E$2 intensities,
the interference effect between the terms of rank even and odd
when the direction of the
magnetic field is reversed, and the
signs of the difference spectrum $\Delta I(\omega)$ around the $E$1 and $E$2 
regions.

When we focus on the detail of the spectra, the whole shape of the spectrum is roughly approximated
by $I_{\textrm{av}}=|\alpha_{1}^{(2)}(\omega) Z_{1}^{(2)}
+\alpha_{2}^{(2)}(\omega) Z_{2}^{(2)}|^2$, i.e., contributions from
the AFQ order parameter are dominant.  
Even around the $E$2 region, about half of the
intensity is supplied by the tail part of the $E$1 signal.
On the other hand, the difference spectrum
$\Delta I(\omega)$ is a direct consequence of the cross
terms between the primary AFQ order and the magnetic-induced
secondary order parameters with odd rank.
A finite intensity of $\Delta I(\omega)$ around the $E$1 peak,
$\Delta I_{E1E1}(\omega)$,
is observed by both the experiment and our calculation.
This term is absent within the fast collision approximation.
The $\Delta I(\omega)$ around the $E$2 position consists
of $\Delta I_{E1E2}(\omega)+\Delta I_{E2E2}(\omega)$.
Comparing the $\Delta I(\omega)$ with the fitting curve based on the 
fast collision approximation constructed by two Lorentzian curves in 
Ref. 31, we see that both give relatively similar outlooks. 
However, this is a coincidence. Our analysis have showed the 
main ingredients of the entire spectrum are four profiles 
$\alpha_{2}^{(1)}(\omega)$, $\alpha_{2}^{(2)}(\omega)$, 
$\alpha_{1}^{(1)}(\omega)$, and $\alpha_{3}^{(2)}(\omega)$. 
We suggest that the difference of the spectral shape between our
calculation and the outcome of the fast collision approximation 
may be observed when the orientation of the applied field is 
changed since the mixing ratios of four profiles are modified. 
A research toward this direction will be a future work.

We also have found that the RXS signal is strong enough
to be detected experimentally at the $L_{2}$ edge,
though the $E$2 peak is practically absent. 
We assert that the spectral analysis based on the microscopic calculation,
which is beyond mere symmetrical consideration,
is very useful to get deeper insights of the RXS phenomena.

As mentioned in the preceding section,
most of our results are robust semi-quantitatively
when several modifications are introduced into the $5d$ DOS.
An exception is the fine structure of $\Delta I(\omega)$.
Utilizing the better (and/or realistic) $5d$ DOS may 
improve the detail of $\Delta I(\omega)$. 
There is a suggestion that the $5d$ DOS of the $e_{g}$ and $t_{2g}$
states are splitting and have the different shapes
derived by a electronic structure calculation.\cite{com.Sakai}
The splitting is expected at least in the 
high temperature region.\cite{Makita08}
We have checked, however, that such difference modifies little the
overall shape of $I_{\textrm{av}}(\omega)$, 
while it affects in a subtle way that of $\Delta I(\omega)$.
If the latter spectrum would be measured with more precision,
the spectral shape can be used to infer the form of the $5d$ DOS.

A recent experimental report on magnetic spin resonance
suggested that the AFQ-based model such as adopted in the
present work met a difficulty in 
explaining its experimental data.\cite{Demishev09}
Even if a tiny amount of antiferromagnetic (AFM) moment is present,
our results survive as long as the moment is small.
A further test to our theory can be performed when the
RXS signals are experimentally examined in the AFM phase.

\begin{acknowledgments}
The authors are grateful to T. Matsumura,
R. Shiina, and O. Sakai for valuable discussions.
This work was partly supported by Grant-in-Aid 
for Scientific Research from the Ministry of Education, Culture, Sport, 
Science, and Technology, Japan.

\end{acknowledgments}

\bibliography{CeB6}

\begin{thebibliography}{54}
\bibitem[{com({\natexlab{b}})}]{com.nagao}
\bibinfo{note}{Note that several typos exist in the corresponding definitions
  of Ref. \onlinecite{Nagao06} are corrected in Ref. \onlinecite{Nagao08}.}
\end{thebibliography}

\end{document}